# Large-bipolaron liquids in cuprate superconductors

David Emin

Department of Physics and Astronomy

University of New Mexico

Albuquerque, New Mexico, 87131, USA

Email: emin@unm.edu

Abstract

Uniquely, large-bipolarons' self-trapped holes occupy superoxygens, each comprising four oxygens circumscribed by four coppers in a $CuO_2$ plane, formed as oxygens relax inward and coppers relax outward. Critically, concomitant oxygen-to-copper electron transfer eliminates copper spins. The *d*-symmetry of superoxygen's groundstate molecular orbital tracks the superoxygens' predominant zero-point radial vibrations. These large bipolarons' distinctive charge transport, absorption, magnetism, local atomic vibrations, condensation into a liquid and subsequent superconductivity are consistent with cuprate superconductors' long-established unusual properties.

Charge carriers in transition-metal oxides are well-known to self-trap thereby forming polarons. Polaron formation results from electronic charge carriers' interactions with displaceable atoms. A stationary charge carrier shifts surrounding atoms' equilibrium positions producing a potential well within which the electronic carrier may be bound. Self-trapping occurs when the rate at which the carrier "circulates" in its bound state (its binding energy divided by Planck's constant) exceeds the displaced atoms' vibrational frequencies. A polaron refers to a self-trapped electronic carrier taken together with the associated altered vibrational state (e.g., shifts of atoms' equilibrium positions and vibrations).[1] The "strong-coupling" polarons and bipolarons discussed here should not be confused with "weak-coupling" polarons whose electronic charge carriers are not self-trapped.[1]

A charge carrier in a covalent material only interacts with the atoms directly linked to the local state it occupies. By contrast, a charge carrier in ionic (polar) materials also interacts at long-range with their displaceable ions. The strength of this long-range (Fröhlich) interaction is proportional to the difference between the reciprocals of a material's high-frequency and static dielectric constants: $(1/\varepsilon_\infty) − (1/\varepsilon_0)$.

Two distinct types of polaron can form in multi-dimensional systems.[1,2] A large polaron's self-trapped electronic carrier extends over multiple structural units. Large polarons can form in ionic solids with strong enough long-range electron-phonon interactions. A small polaron forms when short-range interactions drive its self-trapped electronic carrier to collapse into a single structural unit (e.g., bond, ion, or molecule).

A simple bipolaron is defined as a singlet pair of charge carriers in a common self-trapped orbital. With only a long-range electron-phonon interaction the adiabatic energy of a simple bipolaron is unstable with respect to separating into two polarons.[1,3,4] However, a strong enough short-range electron-phonon interaction will stabilize such a bipolaron. Consider a singlet pair in an embedded plane with electronic bandwidth *T* and short-range electron-lattice interaction $V_S$. Its large bipolaron will be stabilized if $2V_S/T$ is large enough while not so big that it drives collapse into a small bipolaron:

$[4(\varepsilon_0/2\varepsilon_\infty) - 3]/[2(\varepsilon_0/2\varepsilon_\infty)^2 - 1] \leq 2V_S/T < 1$.[1,3,4] Thus, a stable large-bipolaron, rather than a small bipolaron, only becomes possible as the ratio of the material's static to high-frequency dielectric constants rises above 2. Such materials have exceptionally displaceable ions. Examples are found in perovskites and related structures: the insulating parents of superconducting doped $SrTiO_3$ and cuprate superconductors.[5-10]

Features of our strong-coupling large bipolaron are now contrasted with those of an unphysical toy model. Consider self-trapped carriers confined to a plane that is surrounded by displaceable ions (e.g., a $CuO_2$ layer). Equations (7.1) and (7.2) of Ref. 1 are used to obtain the ratio of the radius of a planar singlet large-bipolaron radius to that of a planar large-polaron:

$$\frac{R_{bp}}{R_p} = \frac{(1 - \varepsilon_\infty/\varepsilon_0)(1 - 2V_S/T)}{(1 - 2\varepsilon_\infty/\varepsilon_0)(1 - V_S/T)}. \quad (1)$$

The ratio of the corresponding large-bipolaron binding energy to that for two separated polarons is

$$\frac{E_{bp}}{2E_p} = \left[\frac{(1 - 2\varepsilon_\infty/\varepsilon_0)}{(1 - \varepsilon_\infty/\varepsilon_0)}\right]^2 \frac{(1 - V_S/T)}{(1 - 2V_S/T)}. \quad (2)$$

The conditions $2\varepsilon_\infty/\varepsilon_0 < 1$ and $2V_S/T < 1$, respectively ensure 1) that the pair of self-trapped carriers is bound and 2) that the resulting large bipolaron does not collapse into a small bipolaron. A critical distinctive feature of cuprate superconductors is that $2\varepsilon_\infty/\varepsilon_0 \ll 1$, their ions are exceptionally displaceable. Then Eqs. (1) and (2) indicate that the bipolaron is compact, $R_{bp} < R_p$, and stable with respect to separating into two planar polarons, $-E_{bp} < -2E_p$. Moreover, ubiquitous short-range components of electron-phonon interactions (e.g., deformational potentials) are generally quite strong.[11,12] Therefore, $2V_S$ can approach $T$ to generate a very compact strongly bound large-bipolaron. Nonetheless, toy models for large-bipolaron formation often take $V_S = 0$. Then, $R_{bp} > R_p$ with its simple large bipolaron being unstable with respect to separating into two polarons, $-E_{bp} > -2E_p$. Indeed, for conventional ionic solids (e.g., NaCl), where $2\varepsilon_\infty \approx \varepsilon_0$, these toy large-bipolarons become huge. They only become stabilized with respect to decomposing by including the modest effects of electron correlation.[13] All told, the large bipolarons anticipated in cuprate superconductors are qualitatively different from those studied in unphysical toy models. Henceforth this paper posits the planar area of a large singlet bipolaron in cuprates to be not much larger than a $CuO_2$ unit.

Polaron and bipolaron motion are very slow since they require shifts of atoms' equilibrium position s.[1] Small-(bi)polarons' motions are incoherent (via phonon-assisted hopping).[1] By contrast, large-(bi)polarons' motions are coherent albeit with huge effective masses (e.g., several hundred free-electron masses).[1,14,15] For example, the effective mass of a one-dimensional large polaron of binding energy $E_{aLP}$ produced with acoustic phonons is $m_{aLP} = 4E_{aLP}/s^2$, where $s$ represents the sound velocity.[14]

In transition-metal oxides small-polaron hopping is generally between $d$-states of transition-metal cations and large-polaron's coherent motion is among $p$-states of oxygen anions. Electronic transport measurements indicate 1) small-polaron hopping conduction in p-type doped MnO and 2) large-polaron conduction in p-type doped NiO and CoO and n-type anatase $TiO_2$.[16-18] Coherent electronic transport with gigantic effective masses are reported for putatively doped $La_2CuO_4$.[8]

Large (bi)polarons' self-trapped electronic carriers' relaxation in response to the vibrations of associated atoms lowers their frequencies.[1,14,15] Large (bi)polarons' vibrationally softened regions "reflect" incident ambient phonons.[1,14,15] These phonons' small momenta compared with that of a massive large (bi)polaron results in its very weak scattering. Indeed, a large (bi)polaron's scattering time $\tau$ is proportional to its huge effective mass $m_{LP}$. As a result, the typical large-(bi)polaron mobility is $\mu \equiv q\tau/m_{LP} = qvR^2/kT$, where $v$ denotes a vibrational frequency with $q$ and $R$ representing the self-trapped carrier's charge and radius, respectively. This mobility, only about 1-10 cm$^2$/V-s at room temperature, is comparable to the room-temperature mobility deduced for Sr$_x$La$_{2-x}$CuO$_4$, about 3 cm$^2$/V-s.[19]

Distinctively, the room-temperature mobilities of large-(bi)polarons and those measured in cuprates are much smaller than the minimum mobility for conventional electronic charge carriers.[20] This minimum mobility is reached when a charge carrier's mean-free-path falls to equal its diameter.[1] It is $q\hbar/m_e kT$, 45 cm$^2$/V-s at room temperature when the electronic carrier's effective mass $m_e$ equals the free-electron mass.

Large (bi)polarons' absorption spectra are qualitatively different from those of conventional charge carriers.[21] The Drude fall-off of a large-(bi)polaron's frequency-dependent mobility occurs when the applied frequency exceeds its scattering rate $1/\tau$. A large (bi)polaron's scattering rate is so small that its Drude fall-off occurs at applied frequencies well below those of an atom's vibration: $1/\tau \cong kT/m_{LP}vR^2 \cong v(kT/E_{LP})$, where the large-(bi)polaron binding energy $E_{LP} \gg kT$. Furthermore, these fall-offs shift to lower applied frequencies as the temperature is lowered.

The photon-induced liberation of large-(bi)polarons' self-trapped electronic carriers generates very broad nearly temperature-independent asymmetric absorption bands that emerge at frequencies (> 3$E_{LP}/h$) above those of atomic vibrations.[21] Large-(bi)polaron absorptions fall most slowly with energy beyond those of their peaks. By contrast, small-polarons' broad absorptions fall most rapidly with energy above those of their peaks. Thus, the asymmetries of polarons' high-energy absorption bands distinguish between those from large-(bi)polarons and from small polarons. The absorption spectra generated by super-bandgap excitation of the insulating parents of cuprate superconductors and those of cuprate superconductors are consistent with those expected of large-(bi)polarons.[22-26]

As illustrated in Fig. 6 of Ref. 35, large-(bi)polarons' normal-state absorption spectra have gaps between their low-energy Drude absorptions and their higher-energy excitations of their self-trapped electronic carriers.[21] These "pseudo" energy gaps are not to be mistaken for superconductivity's energy gaps.

Large-(bi)polarons' softened regions can generate local modes in a solid's vibrational spectra.[27] Such local modes are distinguished from a solid's usual atomic vibrations by their lower frequencies and their only occurring at wavelengths that are less than softened regions' linear dimensions, e.g., a large (bi)polaron's diameter. Such "ghost" modes have been observed in addition to among cuprates' intrinsic vibrations.[28] In addition, the localized electronic charge associated with bipolarons annihilates incident positrons, shortening their lifetime.[29] These localized vibrations and charge distributions also disrupt the channeling of injected ions.[30] Strikingly, consistent with increasing carrier homogeneity, their induced shortening of positron lifetimes and disruption of ion channeling progressively disappear as the temperature is lowered below cuprates' superconducting transition temperatures.[29,30]

Large-bipolarons' softened regions also can drive their condensation into a liquid.[31,32] Reducing large-bipolarons' mutual separations below a host material's phonon wavelengths lowers their zero-point vibrational energies. This coherence effect generates intermediate-range phonon-mediated attractions between large bipolarons. These attractions are opposed by "hard-core" repulsions between singlet large-bipolarons which preclude their merger into grander large polarons. These hard-core repulsions occur because the Pauli principle limits occupation of a self-trapping potential well's lowest-energy state to one singlet pair of electronic carriers. The huge static dielectric constants required for large-bipolaron formation can be large enough to suppress slow-moving large-bipolarons' mutual direct Coulomb repulsions. Then the combination of intermediate-range attractions and hard-core repulsions drive the condensation of a large-bipolaron gas to a large-bipolaron liquid. Experiments find that cuprates' normal-state carriers' phase-separate with some coalescing into a liquid.[33,34]

I take large-bipolarons' superconductivity to be a property of its liquid's groundstate. The groundstate remains liquid if its bipolarons do not solidify by ordering commensurate with the underlying solid. A large-bipolaron-liquid's groundstate becomes occupied upon its undergoing a Bose-Einstein condensation. This occurs for charged bosons when their thermal energy becomes comparable to their plasma energy.[35] The large-bipolaron plasma energy is inversely proportional to the square root of the product of the huge large-bipolaron effective mass and its host material's exceptionally large static dielectric constant. As a result, the large-bipolaron plasma energy is that of a typical optic phonon reduced by the factor $(\varepsilon_\infty/\varepsilon_0)1/2$.

A superoxygen is a molecular unit comprising contiguous oxygen atoms.[36] Figure 1 schematically illustrates a superoxygen forms in a $CuO_2$ plane as four oxygen anions circumscribed by four copper cations relax inward in response to the addition of two holes. A large bipolaron in a $CuO_2$ plane extends over a few neighboring superoxygens. Nonetheless, a critical feature of large-bipolaron formation is illustrated by examining just a single superoxygen. Namely, electrons are transferred from the four coalescing oxygen anions to each of the four departing copper cations. This electron transfer occurs because the number of electrons attracted to an oxygen atom falls from two to one (its value for an isolated oxygen atom) as surrounding cations move away from it. In particular, an isolated oxygen atom has a one-electron affinity of 1.5 eV while a second electron is repelled by 7.7. eV.[37] Thus, a valence of 2- is only appropriate for oxygen intimately surrounded by cations. All told, bipolaron formation leaves a singlet electron pair in a non-degenerate four-oxygen molecular orbital as spins are removed from four copper cations: 2 holes + $4O^{2-}$ + $4Cu^{+2}$ → $(O_4)^{2-}$ + $4Cu^{+1}$. Large-bipolaron formation's replacements of four spin-1/2 $Cu^{+2}$ cations with four spinless $Cu^{+1}$ cations produce the observed destruction of the $La_2CuO_4$ magnetism.[38] There is a fundamental difference in how $Cu^{2+}$ spins are affected by paired holes on $O_4$ superoxygens and by holes on oxygen anions forming Zhang-Rice singlets. A Zhang-Rice singlet is formed by anti-ferromagnetically aligning a spin-½ on a $Cu^{2+}$ with a spin-½ carrier on a neighboring $O^{1-}$ whereas electron transfers from a $O_4$ superoxygen to surrounding spin-½ $Cu^{2+}$ cations convert them to spinless $Cu^{1+}$ cations.[39]

Eliminating the entropy of four $Cu^{+2}$ spins also lowers large-bipolarons' Seebeck coefficient, the entropy change upon adding a carrier divided by its electronic charge.[1,15,40] In the high-temperature paramagnetic limit this contribution becomes -$k$ [4 ln(2)]/2$e$ = -120 μV/K while the normal contribution approaches $k$ ln[(1-$c$)/$c$]/2$e$ = 95 μV/K, where 2$e$ denotes the hole-bipolaron-charge and its concentration $c$ is chosen equal to 0.1. Thus, the Seebeck coefficients of large-bipolarons in cuprates differ in magnitude, temperature dependence, and potentially even sign from those of superconducting metals' normal state. The unusual Seebeck coefficients that are observed in cuprates also are compatible with these expectations for large bipolarons.[41-44]

As already noted, distinctive manifestations of large (bi)polarons (mobilities below the minimum for conventional charge carriers, absorption spectra with pseudo gaps, ghost modes, normal-state carriers coalescing into a liquid, carrier-induced loss of magnetism, and unusual Seebeck coefficients) are reported in high-temperature superconductors. Superconductivity in the doped spin-1/2 antiferromagnetic insulator $La_2CuO_4$ is observed over a limited range of dopant concentrations. The concentration of dopants must be large enough for charge carriers to be released from them. The concentration of large-bipolarons also must remain small enough so that their formation is not precluded by interference of their associated ionic displacements. Superconductivity is observed in $Sr_xLa_{2-x}CuO_4$ between doping levels of 5-8% and near 22%. Within this doping range, normal-state resistivities are proportional to temperature.[45] This distinctive feature is consistent with temperature-independent densities of large-(bi) polarons having mobilities proportional to $1/T$. Above the superconductivity's doping regime, charge transport becomes that of a conventional metal.[46]

Large-bipolaron superconductivity should be lost when the groundstate of paired holes order in a manner commensurate with conducting regions of the solid's underlying lattice.[31,32] Ideally, for holes introduced into the square $CuO_2$ planes of tetragonal $Ba_xLa_{2-x}CuO_4$ this solidification should occur when the ratios of the number of holes to the number of $CuO_2$ units are 2/(5×5) = 8%, 2/(4×4) = 1/8 = 12.5% and 2/(3×3) = 22%. Most striking, superconductivity is observed to disappear within the midst of the broad 8%-22% domain, when the doping level is 1/8.[47]

This model envisions a large-bipolaron's self-trapped singlet pair of electrons sharing an oxygen-related molecular orbital.[35] The core of this large-bipolaron is the singlet pair's inwardly shifted anion unit which is surrounded by outwardly shifted heavier copper cations. Four optic-type normal modes describe radial vibrations of these four anions about their displaced equilibrium positions.[35] Interactions between neighboring oxygens generate their vibrational dispersion. The lowest-frequency normal mode has the largest amplitude. As schematically depicted in Fig. 2, this dominant mode has each of the four oxygens vibrating in opposition to its two neighboring oxygens. The molecular orbital of this large-bipolarons' self-trapped electron-pair garners the dynamic $d_{xy}$-symmetry of the four oxygens' radial vibrations.[35] As the four oxygens of the $(O_4)^{2-}$ molecular orbital radially vibrate, its two added electrons adiabatically transfer from the inward moving oxygens to those moving outwardly toward the surrounding copper cations. The superconducting state retains this anisotropy since its large-bipolaron-liquid's very formation is driven by its self-trapped pairs' coherent response to long-wavelength zero-point phonons. This result is consistent with the measured anisotropy in the $a-b$ plane of cuprates' superconductivity.[48]

In summary, cuprates' unusual normal-state transport indicates large bipolarons. Gigantic effective masses characterize a large-bipolaron's very slow coherent motion since it requires substantial shifts of surrounding atoms' equilibrium positions. Large bipolarons' huge thermal momenta ensure that they are only weakly scattered by typical phonons. Large-bipolarons' extraordinarily long scattering times also limit measurement of their mobilities to applied frequencies below characteristic phonon frequencies. Increasing scattering time upon cooling widens the "pseudo" gap between large-bipolarons' very low-frequency (Drude) absorption and that from exciting its self-trapped carriers. Distinctively, room-temperature large-bipolaron mobilities are much smaller than those permitted of conventional carriers. Large-bipolaron formation increases equilibrium separations between associated anions and cations thereby reducing the charge transfer between them. As schematically illustrated in Fig. 1, the core of a cuprate planar large-bipolaron is modelled as a non-magnetic superoxide dianion $(O_4)^{2-}$ surrounded by four non-magnetic $Cu^{1+}$ cations. Large-bipolarons' elimination of spin entropy generates a contribution to the Seebeck coefficient which grows increasingly negative with rising temperature. By contrast, the Seebeck coefficient of a normal p-type metal becomes increasingly positive with rising temperature. These features explain cuprates' unusual normal-state transport: exceptionally low mobilities, resistivities nearly proportional to temperature, temperature-dependent "pseudo" gaps and unusual Seebeck coefficients.

Distinctive features of cuprate superconducting condensate also indicate large-bipolarons. The vibrations of atoms whose equilibrium positions are shifted by large-bipolaron formation are softened by its self-trapped carriers' relaxation. These locally softened vibrations generate "ghost modes" that appear only at phonon wavelengths shorter than large-bipolarons' spatial extents. In addition, vibrational energies are lowered as the separations between large-bipolarons are reduced below other phonons' long wavelengths. This coherence effect can drive the phonon-assisted condensation of large bipolarons into a liquid. This large-bipolaron liquid will superconduct upon condensing to a fluid groundstate. The condensation temperature is comparable to the characteristic phonon temperature reduced by $(\varepsilon_\infty/\varepsilon_0)1/2$. However, superconductivity will be lost if the large-bipolarons order commensurate with the underlying $La_2CuO_4$ lattice. Indeed, superconductivity of single-phase tetragonal $Ba_xLa_{2-x}CuO_4$ disappears for x = 2/16, corresponding to large bipolarons solidifying into a 4x4 square superlattice. As schematically illustrated in Fig. 2, the lowest-energy largest-amplitude zero-point "radial" vibration of the displaced atoms of a four-fold planar large bipolaron has $d_{xy}$ symmetry. Collective motions of the condensate's atoms and the associated electrons will also manifest this symmetry.

All told, well-known unusual normal-state and superconducting properties of cuprate superconductors are consistent with their charge carriers being oxygen-related large bipolarons.

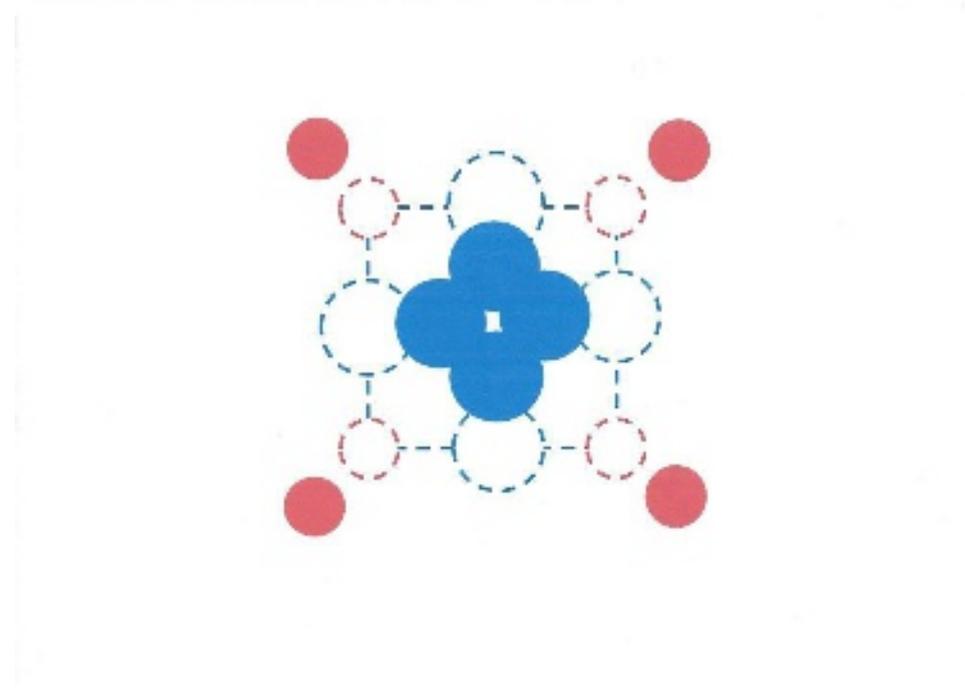

Fig. 1 A superoxygen dianion forms as a pair of holes added to oxygen anions (blue circles) of a CuO$_2$ unit induces them to move inward as copper cations (red circles) move outward. Concomitant transfers of electrons to these copper cations eliminate their spins.

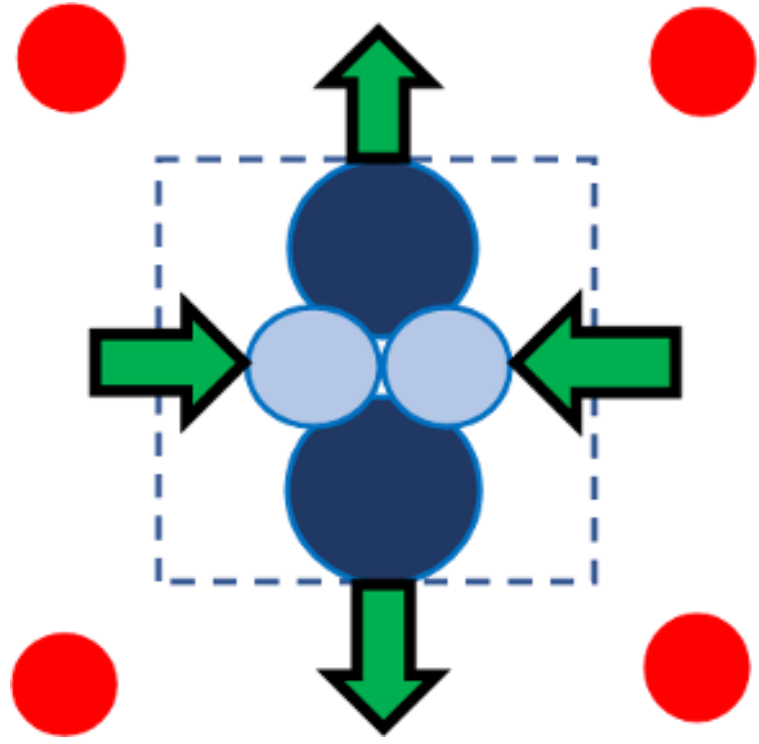

Fig. 2 Arrows depict the relative movements of oxygen-anions' for the lowest-frequency and largest-zero-point-amplitude normal radial vibrational mode. The $(O_4)^{2-}$ molecular orbital's two added electrons adiabatically transfer from inward moving oxygens (lightened blue circles) to outward moving oxygens (darkened blue circles) that are closer to the surrounding copper cations.